\documentclass[conference]{IEEEtran}
\usepackage{amsmath}
\usepackage{amssymb}
\usepackage[dvips]{graphicx}
\usepackage{epsfig}

\newcommand{\X}{\mathbf{X}}

\newcommand{\hh}{\mathbf{H}}
\newcommand{\w}{\mathbf{w}}

\newcommand{\I}{\mathbf{I}}

\newcommand{\bv}{\mathbf{v}}

\newcommand{\bu}{\mathbf{u}}

\newtheorem{lemma:bitenergylow}{Lemma}
\newtheorem{lemma:bitenergyhigh}[lemma:bitenergylow]{Lemma}

\newtheorem{prop:asympcap}{Theorem}
\newtheorem{prop:flashminbitenergy}[prop:asympcap]{Theorem}
\newtheorem{prop:flashbitenergy}[prop:asympcap]{Theorem}
\newtheorem{prop:pasympcap}[prop:asympcap]{Theorem}

\begin{document}


\title{Secure Relay Beamforming over Cognitive Radio Channels}



%
\author{\authorblockN{Junwei Zhang and Mustafa Cenk Gursoy}
\authorblockA{Department of Electrical Engineering\\
University of Nebraska-Lincoln, Lincoln, NE 68588\\ Email:
junwei.zhang@huskers.unl.edu, gursoy@engr.unl.edu}}

\maketitle
\begin{abstract}\footnote{This work was supported by the National Science Foundation under Grants CNS--0834753, and CCF--0917265.}
In this paper, a cognitive relay channel is considered, and amplify-and-forward (AF) relay beamforming designs in
the presence of an eavesdropper and a primary
user are studied. Our objective is to optimize the performance of
the cognitive relay beamforming system while limiting the
interference in the direction of the primary receiver and keeping the transmitted signal secret from the eavesdropper. We show that under
both total and individual power constraints, the problem becomes a
quasiconvex optimization problem  which can be solved by interior
point  methods. We also propose two sub-optimal null space
beamforming schemes which are obtained in a more computationally efficient way.

\end{abstract}

\section{introduction}
The need for the efficient use of the scarce spectrum in wireless
applications has led to significant interest in the analysis of
cognitive radio systems. One possible scheme for the operation of the cognitive
radio network is to allow the secondary users to transmit
concurrently on the same frequency band with the primary users as
long as the resulting interference power at the primary receivers is
kept below the interference temperature limit \cite{Haykin}. Note that interference to the primary users is caused due to the broadcast nature of wireless transmissions, which allows
the signals to be received by all users within the communication
range. Note further that this broadcast nature also makes wireless communications vulnerable to eavesdropping.
The problem of secure transmission in the presence of an
eavesdropper was first studied from an information-theoretic
perspective in \cite{wyner} where Wyner considered a wiretap channel
model. In \cite{wyner}, the secrecy capacity is defined as the maximum
achievable rate from the transmitter to the legitimate receiver,
which can be attained  while keeping the eavesdropper completely
ignorant of the transmitted messages. Later, Wyner's result was extended to the Gaussian channel in
\cite{cheong}. Recently, motivated by the importance of security in wireless applications, information-theoretic security has been investigated in fading multi-antenna and multiuser channels. For instance, cooperative relaying under secrecy
constraints was studied in
\cite{dong}--\cite{jzhang1}. In \cite{jzhang1},
for amplify and forwad relaying scheme, not having analytical solutions for the optimal beamforming design under both total and individual power constraints, an iterative algorithm is proposed to numerically obtain the optimal beamforming structure and maximize the secrecy rates.

Although cognitive radio networks are also susceptible to
eavesdropping, the combination of cognitive radio channels and
information-theoretic security has received little attention. Very recently, Pei \emph{et al.} in \cite{Pei} studied secure communication
over multiple input, single output (MISO) cognitive radio channels. In this work,  finding the secrecy-capacity-achieving transmit covariance matrix
under joint transmit and interference power constraints is formulated as a quasiconvex optimization
problem.

In this paper, we investigate the collaborative relay beamforming
under secrecy constraints in the cognitive radio network. We first characterize
the secrecy rate of the amplify-and-forward (AF) cognitive relay
channel. Then, we formulate the beamforming optimization as a quasiconvex optimization
problem which can be solved through convex semidefinite
programming (SDP). Furthermore, we propose two sub-optimal null space beamforming schemes to reduce the computational complexity.

\section{Channel Model}

\begin{figure}
\begin{center}
\includegraphics[width = 0.5\textwidth]{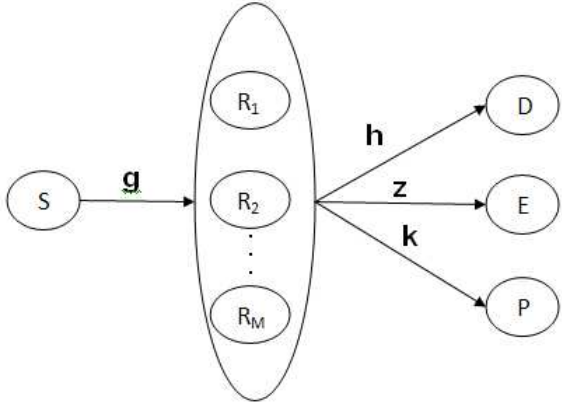}
\caption{Channel Model} \label{fig:channel}
\end{center}
\end{figure}

We consider a cognitive relay
channel with a secondary user source $S$, a primary user $P$,
a secondary user destination $D$, an eavesdropper $E$,  and $M$
relays $\{R_m\}_{m=1}^M$, as depicted in Figure \ref{fig:channel}. We
assume that there is no direct link between $S$ and $D$, $S$ and
$P$, and $S$ and $E$. We also assume that relays work synchronously to perform beamforming
by multiplying the signals to be transmitted with complex weights
$\{w_m\}$. We denote the channel fading
coefficient between $S$ and $R_m$ by $g_m\in \mathbb{C}$, the fading
coefficient between $R_m$ and $D$ by $h_m\in \mathbb{C}$, $R_m$ and
$P$ by $k_m\in \mathbb{C}$ and the fading coefficient between $R_m$
and $E$ by $z_m\in \mathbb{C}$. In this model, the source $S$ tries
to transmit confidential messages to $D$ with the help of the relays on the same band as the primary user's
while keeping the interference on the
primary user below some predefined interference temperature limit and keeping the eavesdropper
$E$ ignorant of the information.
It's obvious that our channel is a two-hop relay network. In the
first hop, the source $S$ transmits $x_s$ to relays with power
$E[|x_s|^2]=P_s$. The received signal at the $m^{\text{th}}$ relay  $R_m$ is given by
\begin{align}
y_{r,m}=g_m x_s+\eta_m
\end{align}
where $\eta_m$ is the background noise that has a Gaussian
distribution with zero mean and variance of $N_m$.

In the AF scenario, the received
signal at $R_m$ is directly multiplied by
$l_mw_m$ without decoding, and forwarded to $D$. The relay output
can be written as
\begin{align}
x_{r,m}=w_m l_m (g_m x_s+ \eta_m).
\end{align}
The scaling factor,
\begin{align}
l_m=\frac{1}{\sqrt{|g_m|^2P_s+N_m}},
\end{align}
is used to ensure $E[|x_{r,m}|^2]=|w_m|^2$. There are two kinds of
power constraints for relays. First one is  a total relay power
constraint in the following form: $||\w||^2=\w^\dagger \w\leq P_T$
where $\w=[w_1,...w_M]^T$ and $P_T$ is the maximum total power. $(\cdot)^T$ and $(\cdot)^\dagger $ denote the transpose
and conjugate transpose, respectively, of a matrix or vector. In a
multiuser network such as the relay system we study in this paper,
it is practically more relevant to consider individual power
constraints as wireless nodes generally operate under such
limitations. Motivated by this, we can impose $|w_m|^2 \leq p_m
\forall m $ or equivalently $|\w|^2 \leq \mathbf{p}$ where
$|\cdot|^2$ denotes the element-wise norm-square operation and
$\mathbf{p}$ is a column vector that contains the components
$\{p_m\}$. $p_m$ is the maximum power for the $m^{\text{th}}$ relay node.

The
received signals at the destination $D$ and eavesdropper $E$ are the
superposition of the messages sent by the relays. These received
signals are expressed, respectively, as
\begin{align}
y_d&=\sum_{m=1}^M h_m w_m l_m (g_m x_s +\eta_m) +n_0, ~\text{and} \\
 y_e&=\sum_{m=1}^M z_m w_m l_m (g_m x_s+\eta_m) +n_1
\end{align}
where $n_0$ and $n_1$ are the Gaussian background noise
components with zero mean and
variance $N_0$, at $D$  and $E$, respectively. It is easy to compute  the received SNR at $D$
and $E$ as
\begin{align}
\Gamma_d&=\frac{|\sum_{m=1}^M h_m g_m l_m w_m|^2 P_s}{\sum_{m=1}^M |h_m|^2l_m^2 |w_m|^2 N_m +N_0}, ~\text{and} \\
\Gamma_e&=\frac{|\sum_{m=1}^M z_m g_m l_m w_m|^2 P_s}{\sum_{m=1}^M
|z_m|^2l_m^2 |w_m|^2 N_m +N_0}.
\end{align}
The secrecy rate is now given by
\begin{align}
R_s&=I(x_s;y_d)-I(x_s;y_e)\\
&=\log(1+\Gamma_d)-\log(1+\Gamma_e)\\
&=\log\Bigg(\frac{\sum_{m=1}^M |z_m|^2l_m^2 |w_m|^2 N_m
+N_0}{\sum_{m=1}^M |h_m|^2l_m^2 |w_m|^2 N_m +N_0}\times \nonumber \\
&\hspace{-.4cm}\frac{|\sum_{m=1}^M h_m g_m l_m w_m|^2 P_s+\sum_{m=1}^M
|h_m|^2l_m^2 |w_m|^2 N_m +N_0}{|\sum_{m=1}^M z_m g_m l_m w_m|^2
P_s+\sum_{m=1}^M |z_m|^2l_m^2 |w_m|^2 N_m +N_0}\Bigg)\label{srate}
\end{align}
where $I(\cdot;\cdot)$ denotes the mutual information. The
interference at the primary user is
\begin{align}
\Lambda=|\sum_{m=1}^M k_m g_m l_m w_m|^2 P_s+\sum_{m=1}^M
|k_m|^2l_m^2 |w_m|^2 N_m.
\end{align}
In this paper, under the assumption that the relays have perfect channel side information (CSI), we address the joint optimization of $\{w_m\}$ and hence identify the optimum collaborative
relay beamforming (CRB) direction that maximizes the secrecy rate in
(\ref{srate}) while maintaining the interference on the primary user
under a certain threshold, i.e,. $\Lambda\leq \gamma$, where $\gamma$ is the
interference temperature limit.

\section{Optimal  Beamforming}\label{sec:op}
 Let us define
\begin{align}
\mathbf{h_g}&=[h_1^* g_1^* l_1,... ,h_M^* g_M^* l_M]^T, \\
\mathbf{h_z}&=[z_1^* g_1^* l_1,... ,z_M^* g_M^* l_M]^T, \\
\mathbf{h_k}&=[k_1^* g_1^* l_1,... ,k_M^* g_M^* l_M]^T, \\
\mathbf{D_h}&=\text{Diag}(|h_1|^2l_1^2N_1,...,|h_M|^2l_M^2N_M),  \\
\mathbf{D_z}&=\text{Diag}(|z_1|^2l_1^2N_1,...,|z_M|^2l_M^2N_M),~\text{and}\\
\mathbf{D_k}&=\text{Diag}(|k_1|^2l_1^2N_1,...,|k_M|^2l_M^2N_M)
\end{align}
where superscript $*$ denotes conjugate operation. Then, the
received SNR at the destination and eavesdropper, and the
interference on primary user  can be written, respectively, as
\begin{align}
\Gamma_d&=\frac{ P_s  \w^\dagger \mathbf{h_g} \mathbf{h_g}^\dagger
\w }{\w^\dagger \mathbf{D_h} \w  +N_0}, \\
\Gamma_e&=\frac{P_s  \w^\dagger \mathbf{h_z}\mathbf{h_z}^\dagger
\w}{\w^\dagger \mathbf{D_z} \w  +N_0},\\
\Lambda&=P_s  \w^\dagger \mathbf{h_k}\mathbf{h_k}^\dagger
\w+\w^\dagger \mathbf{D_k} \w .
\end{align}
With these notations, we can write the objective function of the
optimization problem (i.e., the term inside the logarithm in (\ref{srate})) as
\begin{align}
&\frac{1+\Gamma_d}{1+\Gamma_e}\nonumber =\frac{1+\frac{ P_s
\w^\dagger \mathbf{h_g} \mathbf{h_g}^\dagger \w }{\w^\dagger
\mathbf{D_h} \w +N_0}}{1+\frac{P_s  \w^\dagger
\mathbf{h_z}\mathbf{h_z}^\dagger \w}{\w^\dagger \mathbf{D_z} \w
+N_0}}\\
&=\frac{\w^\dagger \mathbf{D_h} \w  +N_0+P_s  \w^\dagger
\mathbf{h_g} \mathbf{h_g}^\dagger \w }{\w^\dagger \mathbf{D_z} \w
+N_0+P_s  \w^\dagger \mathbf{h_z}\mathbf{h_z}^\dagger \w} \times
\frac{\w^\dagger \mathbf{D_z} \w +N_0}{\w^\dagger \mathbf{D_h} \w
+N_0}\\
&=\frac{N_0+tr((\mathbf{D_h}+P_s\mathbf{h_g} \mathbf{h_g}^\dagger)\w
\w^\dagger)}{N_0+tr((\mathbf{D_z}+P_s\mathbf{h_z}
\mathbf{h_z}^\dagger)\w \w^\dagger)} \times
\frac{N_0+tr(\mathbf{D_z} \w \w^\dagger)}{N_0+tr(\mathbf{D_h} \w
\w^\dagger)}. \nonumber
\end{align}
If we denote $t_1=\frac{N_0+tr((\mathbf{D_h}+P_s\mathbf{h_g}
\mathbf{h_g}^\dagger)\w
\w^\dagger)}{N_0+tr((\mathbf{D_z}+P_s\mathbf{h_z}
\mathbf{h_z}^\dagger)\w \w^\dagger)}$,
$t_2=\frac{N_0+tr(\mathbf{D_z} \w \w^\dagger)}{N_0+tr(\mathbf{D_h}
\w \w^\dagger)}$, define $\X\ \triangleq \w \w ^\dagger$, and employ the semidefinite relaxation approach, we can
express the beamforming optimization problem as
\begin{align}
\begin{split}\label{optimal}
&\max_{\X, t_1, t_2} ~~~t_1t_2  \\
&s.t ~~ tr\left(\X\left(\mathbf{D_h}+P_s\mathbf{h_g}
\mathbf{h_g}^\dagger- t_1\left(\mathbf{D_z}+P_s\mathbf{h_z}
\mathbf{h_z}^\dagger\right)\right)\right)\geq N_0(t_1-1)\\
 &~~ tr\left(\X\left(\mathbf{D_z}-
t_2\mathbf{D_h}\right)\right)\geq N_0(t_2-1)\\
&~~tr\left(\X\left(\mathbf{D_k}+P_s\mathbf{h_k}
\mathbf{h_k}^\dagger\right)\right)\leq\gamma\\
 &and ~~diag(\X)\leq
\mathbf{p},~~(and/or~~ tr(\X) \leq P_T)~~~ and ~~~\X \succeq 0.
\end{split}
\end{align}
The optimization problem here is similar to that in
\cite{jzhang1}. The only difference is that we have an additional
constraint due to the interference limitation.
Thus, we can use the same optimization framework. The optimal beamforming solution that maximizes the
secrecy rate in the cognitive relay channel can be obtained by
using semidefinite programming with a two dimensional search for both
total and individual power constraints. For simulation, one can use
the well-developed interior point method based package SeDuMi
\cite{sedumi}, which produces a feasibility certificate if the
problem is feasible, and its popular interface Yalmip \cite{yalmip}. It is important to note that
we should have the optimal $\X$ to be of rank-one to determine the
beamforming vector. While proving analytically the existence of a rank-one solution for the above optimization problem seems to be a difficult task\footnote{Since we in general have more than two linear
constraints depending on the number of relay nodes and since we
cannot assume that we have channels with real and positive
coefficients, the techniques that are used in several studies to prove the existence of a rank-one solution
(see e.g., \cite{Gan}, \cite{luo1},and references therein) are not directly
applicable to our setting.}, we would like to emphasize that the solutions
are rank-one in our simulations. Thus, our numerical result are
tight. Also, even in the case we encounter a solution with rank higher than
one, the Gaussian randomization technique is practically proven to
be effective in finding a feasible, rank-one approximate solution of
the original problem.  Details can be found in \cite{luo1}.

\section{Sub-Optimal Null Space Beamforming}\label{sec:null}
Obtaining the optimal solution requires significant computation. To simplify the analysis, we propose
suboptimal null space beamforming techniques in this section .

\subsection{Beamforming in the Null Space of Eavesdropper's Channel (BNE)}
We choose $\w$ to lie in the null space of $\mathbf{h_z}$. With
this assumption, we eliminate $E$'s capability of eavesdropping on
$D$. Mathematically, this is equivalent to $|\sum_{m=1}^M z_m g_m
l_m w_m|^2=|\mathbf{h_z}^\dagger \w|^2=0$, which means $\w$ is in
the null space of $\mathbf{h_z}^\dagger$. We can write
$\w=\hh_{z}^\bot \bv$, where $\hh_{z}^\bot$ denotes the projection
matrix onto the null space of $\mathbf{h_z}^\dagger$. Specifically,
the columns of $\hh_{z}^\bot$ are orthonormal vectors which form the
basis of the null space of $\mathbf{h_z}^\dagger$. In our case,
$\hh_{z}^\bot$ is an $M\times(M-1)$ matrix. The total power
constraint becomes $\w^\dagger \w= \bv^\dagger {\hh_{z}^\bot}^\dagger
\hh_{z}^\bot \bv=\bv^\dagger \bv\leq P_T$. The individual power
constraint becomes $|\hh_{z}^\bot \bv|^2 \leq \mathbf{p}$

Under the above null space beamforming assumption,
$\Gamma_e$ is zero. Hence, we only need to maximize  $\Gamma_d$ to get
the highest achievable secrecy rate. $\Gamma_d$ is now expressed as
\begin{align}
\Gamma_d&=\frac{ P_s  \bv^\dagger {\hh_{z}^\bot}^\dagger\mathbf{h_g}
\mathbf{h_g}^\dagger\hh_{z}^\bot \bv}{\bv^\dagger
{\hh_{z}^\bot}^\dagger \mathbf{D_h} \hh_{z}^\bot\bv  +N_0}.
\end{align}
The interference on the primary user can be written as
\begin{align}
\Lambda&=P_s  \bv^\dagger {\hh_{z}^\bot}^\dagger\mathbf{h_k}
\mathbf{h_k}^\dagger\hh_{z}^\bot \bv+\bv^\dagger
{\hh_{z}^\bot}^\dagger \mathbf{D_k} \hh_{z}^\bot\bv.
\end{align}
Defining $\X\ \triangleq \bv \bv$, we can express the
optimization problem as
\begin{align}
\begin{split}\label{t1}
&\max_{\X, t} ~~~t  \\
&\text{s.t} ~~
\text{tr}\left(\X\left(P_s{\hh_{z}^\bot}^\dagger\mathbf{h_g}
\mathbf{h_g}^\dagger\hh_{z}^\bot-
t{\hh_{z}^\bot}^\dagger \mathbf{D_h} \hh_{z}^\bot\right)\right)\geq N_0t\\
&~~tr\left(\X\left({\hh_{z}^\bot}^\dagger \mathbf{D_k}
\hh_{z}^\bot+P_s{\hh_{z}^\bot}^\dagger\mathbf{h_k}
\mathbf{h_k}^\dagger\hh_{z}^\bot\right)\right)\leq\gamma\\
&\text{and} ~~\text{diag}(\hh_{z}^\bot\X{\hh_{z}^\bot}^\dagger)\leq
\mathbf{p},(and/or~~ tr(\X) \leq P_T)~~~ and ~~~\X \succeq 0.
\end{split}
\end{align}
This problem can be easily solved by semidefinite programming with
bisection search \cite{jzhang}.

\subsection{Beamforming in the Null Space of Eavesdropper's and Primary User's Channels (BNEP)}

In this section, we choose $\w$ to lie in the null space of
$\mathbf{h_z}$ and $\mathbf{h_k}$. Mathematically, this is
equivalent to requiring $|\sum_{m=1}^M z_m g_m l_m
w_m|^2=|\mathbf{h_z}^\dagger \w|^2=0$, and $|\sum_{m=1}^M k_m g_m l_m
w_m|^2=|\mathbf{h_k}^\dagger \w|^2=0$. We can write
$\w=\hh_{z,k}^\bot \bv$, where $\hh_{z,k}^\bot$ denotes the
projection matrix onto the null space of $\mathbf{h_z}^\dagger$ and
$\mathbf{h_k}^\dagger$. Specifically, the columns of $\hh_{z,k}^\bot$
are orthonormal vectors which form the basis of the null space. In
our case, $\hh_{z,k}^\bot$ is an $M\times(M-2)$ matrix. The total
power constraint becomes $\w^\dagger \w= \bv^\dagger
{\hh_{z,k}^\bot}^\dagger \hh_{z,k}^\bot \bv=\bv^\dagger \bv\leq
P_T$. The individual power constraint becomes $|\hh_{z,k}^\bot
\bv|^2 \leq \mathbf{p}$.

With this beamforming strategy, we again have $\Gamma_e = 0$. Moreover, the interference on the primary user is now
reduced to
\begin{align}
\Lambda=\sum_{m=1}^M |k_m|^2l_m^2 |w_m|^2 N_m=\bv^\dagger
{\hh_{z,k}^\bot}^\dagger \mathbf{D_k} \hh_{z,k}^\bot\bv
\end{align}
which is the sum of the forwarded additive noise components present at the relays.
Now, the optimization problem becomes
\begin{align}
\begin{split}\label{t1}
&\max_{\X, t} ~~~t  \\
&\text{s.t} ~~
\text{tr}\left(\X\left(P_s{\hh_{z,k}^\bot}^\dagger\mathbf{h_g}
\mathbf{h_g}^\dagger\hh_{z,k}^\bot-
t{\hh_{z,k}^\bot}^\dagger \mathbf{D_h} \hh_{z,k}^\bot\right)\right)\geq N_0t\\
&~~tr\left(\X\left({\hh_{z,k}^\bot}^\dagger \mathbf{D_k}
\hh_{z,k}^\bot\right)\right)\leq\gamma\\
&\text{and}
~~\text{diag}(\hh_{z,k}^\bot\X{\hh_{z,k}^\bot}^\dagger)\leq
\mathbf{p},(and/or~~ tr(\X) \leq P_T) ~~~ \\ &\text{and} ~~~\X \succeq 0.
\end{split}
\end{align}
Again, this problem can be solved through semidefinite programming. With the following assumptions, we can also obtain a closed-form characterization of the beamforming structure. Since the interference experienced by the primary user consists of the forwarded noise components, we can assume that the interference constraint $\Lambda \leq \gamma$
is inactive unless $\gamma$ is very small. With this assumption, we can drop this constraint. If we further
assume that the relays operate under the total power constraint
expressed as $\bv^\dagger \bv\leq P_T$, we can get the following closed-form
solution:
\begin{align}
\begin{split}\label{eq:maxsecrecyrate}
& \max_{\bv^\dagger \bv\leq P_t} \Gamma_d \\
&= \max_{\bv^\dagger \bv\leq P_t}\frac{ P_s  \bv^\dagger
{\hh_{z,k}^\bot}^\dagger\mathbf{h_g} \mathbf{h_g}^\dagger\hh_{z,k}^\bot
\bv}{\bv^\dagger {\hh_{z,k}^\bot}^\dagger \mathbf{D_h} \hh_{z,k}^\bot\bv
+N_0}\\
&=\max_{\bv^\dagger \bv\leq P_t}\frac{ P_s  \bv^\dagger
{\hh_{z,k}^\bot}^\dagger\mathbf{h_g} \mathbf{h_g}^\dagger\hh_{z,k}^\bot
\bv}{\bv^\dagger \left({\hh_{z,k}^\bot}^\dagger \mathbf{D_h}
\hh_{z,k}^\bot+\frac{N_0}{P_T}\I\right)\bv}\\
&=P_s\lambda_{max}\left( {\hh_{z,k}^\bot}^\dagger\mathbf{h_g}
\mathbf{h_g}^\dagger\hh_{z,k}^\bot ,{\hh_{z,k}^\bot}^\dagger
\mathbf{D_h} \hh_{z,k}^\bot+\frac{N_0}{P_T}\I\right) \nonumber
\end{split}
\end{align}
where $\lambda_{\max}(\mathbf{A},\mathbf{B})$ is the largest
generalized eigenvalue of the matrix pair $(\mathbf{A},\mathbf{B})$
\footnote{For a Hermitian matrix $\mathbf{A} \in \mathbb{C}^{n\times
n}$ and positive definite matrix $\mathbf{B} \in \mathbb{C}^{n\times
n}$, $(\lambda,\psi)$ is referred to as a generalized eigenvalue --
eigenvector pair of $(\mathbf{A},\mathbf{B})$ if $(\lambda,\psi)$
satisfy $\mathbf{A} \psi=\lambda \mathbf{B} \psi$ \cite{matrix}.}.
Hence, the maximum secrecy rate is
achieved by the beamforming vector
$
\bv_{opt}=\varsigma \bu
$
where $\bu$ is the eigenvector that corresponds to
$\lambda_{\max}\left( {\hh_{z}^\bot}^\dagger\mathbf{h_g}
\mathbf{h_g}^\dagger\hh_{z}^\bot,{\hh_{z}^\bot}^\dagger \mathbf{D_h}
\hh_{z}^\bot+\frac{N_0}{P_T}\I\right)$  and $\varsigma$ is chosen to
ensure $\bv_{opt}^\dagger \bv_{opt} =P_T$.

\section{Multiple Primary Users and Eavesdroppers}
The discussion in Section \ref{sec:op} can be
easily extended to the case of more than one primary user  in the
network. Each primary user will introduce an
interference constraint $\Gamma_i\leq\gamma_i$ which can be
straightforwardly included into (\ref{optimal}). The beamforming optimization is still a semidefinite programming problem. On the other hand, the
results in Section \ref{sec:op} cannot be easily extended to the
multiple-eavesdropper scenario. In this case, the secrecy rate for AF
relaying is
$R_s=I(x_s;y_d)-\max_{i} I(x_s;y_{e,i})$, where the maximization is
over the rates achieved over the links between the relays and
different eavesdroppers. Hence, we have to consider the eavesdropper
with the strongest channel. In this scenario, the objective function
cannot be expressed in the form given in (\ref{srate}) and
the optimization framework provided in Section \ref{sec:op} does not directly apply
to the multi-eavesdropper model.

However, the null space beamforming  schemes discussed in
Section \ref{sec:null} can be extended to the case of multiple primary users and
eavesdroppers under the condition that the number of relay nodes is greater than
the number of eavesdroppers or the total number of eavesdroppers and
primary users depending on which null space beamforming is used. The
reason for this condition is to make sure the projection matrix $\hh^\bot$
exists. Note that the null space of $i$ channels in general has the
dimension $M \times (M-i)$ where $M$ is the number of relays.

\section{Numerical Results and Discussion}
We assume  that $\{g_m\}$, $\{h_m\}, \{z_m\}, \{k_m\}$ are complex,
circularly symmetric Gaussian random variables with zero mean and
variances $\sigma_g^2$, $\sigma_h^2$, $\sigma_z^2$ and $\sigma_k^2$
respectively.  In this section, each figure is plotted for fixed
realizations of the Gaussian channel coefficients. Hence, the
secrecy rates in the plots are instantaneous secrecy rates.

\begin{figure}
\begin{center}
\includegraphics[width = 0.5\textwidth]{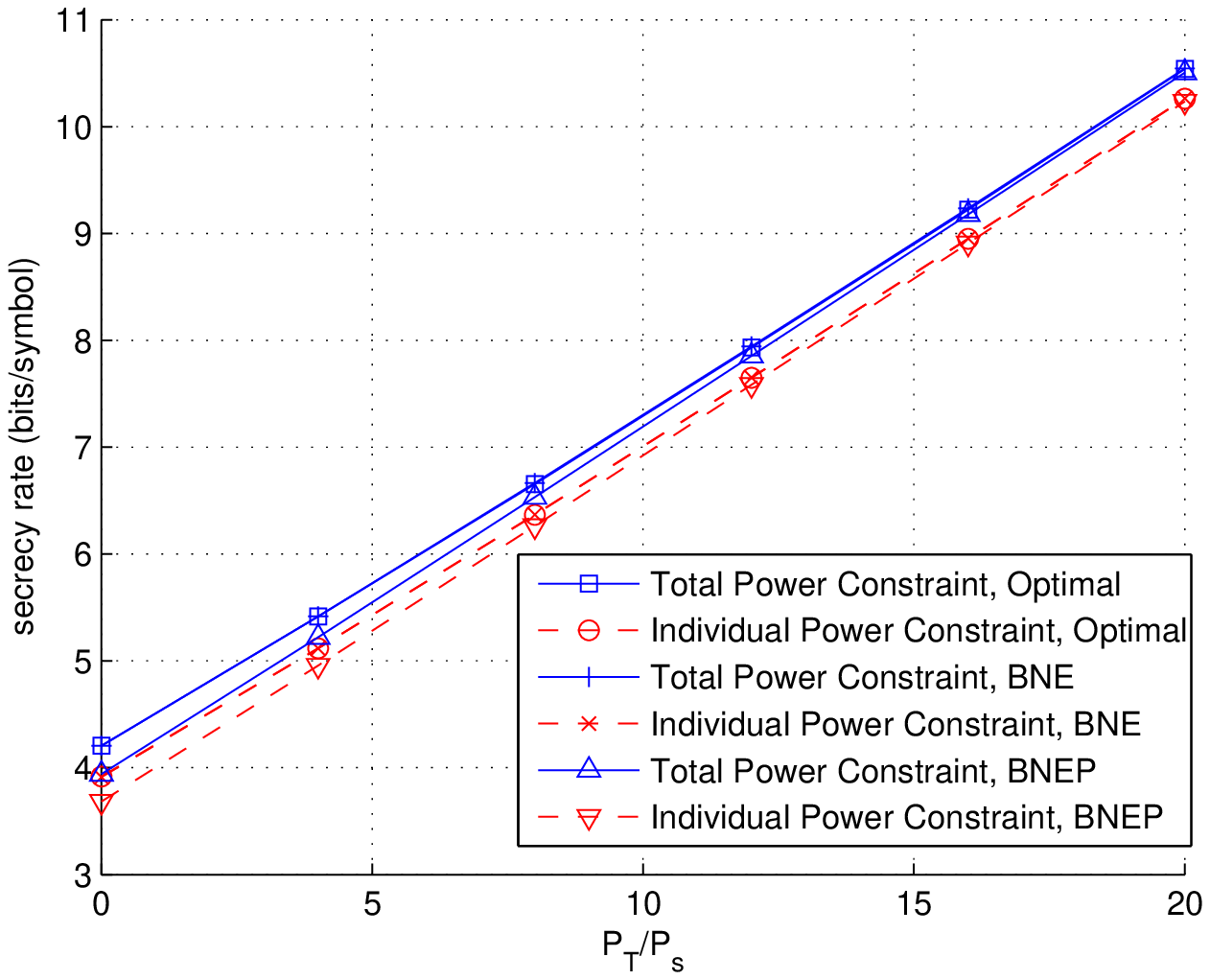}
\caption{AF secrecy rate vs. $P_T/P_s$. $ \sigma_g= 10, \sigma_h=,
\sigma_z=1, \sigma_k=1, M=10, \gamma=0 dB$. } \label{fig:1}
\end{center}
\end{figure}

In Fig. \ref{fig:1}, we plot the optimal secrecy rates for the
amplify-and-forward collaborative relay beamforming system under both
individual and total power constraints. We also provide, for comparison, the secrecy rates attained by using the suboptimal beamforming schemes. The fixed
parameters are $\sigma_g= 10, \sigma_h=1, \sigma_z=1, \sigma_k=1$,
$\gamma=0dB$, and $M=10$. Since AF secrecy rates depend on both
the source and relay powers, the rate curves are plotted as a
function of $P_T/P_s$. We assume that the relays have equal powers
in the case in which individual power constraints are imposed, i.e.,
$p_i = P_T/M$. It is immediately seen from the figure that the
suboptimal null space beamforming achievable rates under both total
and individual power constraints are very close to the corresponding
optimal ones. Especially, they are nearly identical in the high SNR
regime, which suggests that null space beamforming is optimal at high
SNRs. Thus, null space beamforming schemes are good alternatives
as they are obtained with
much less computational burden. Moreover, we interestingly observe
that imposing individual relay power constraints leads to  small
losses in the secrecy rates.

\begin{figure}
\begin{center}
\includegraphics[width = 0.5\textwidth]{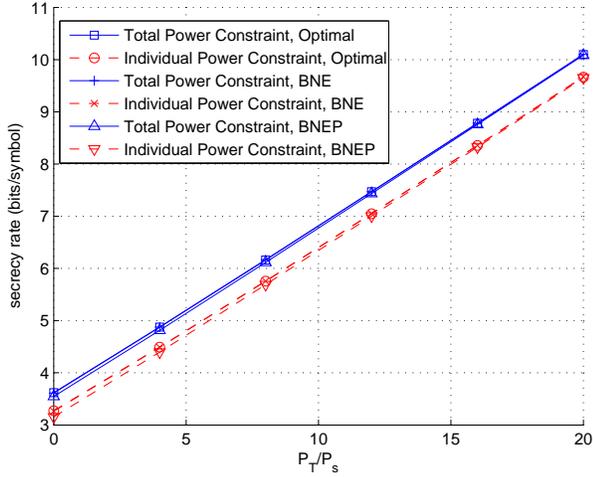}
\caption{AF secrecy rate vs. $P_T/P_s$. $ \sigma_g= 10, \sigma_h=1,
\sigma_z=2, \sigma_k=4, M=10, \gamma=10 dB$. } \label{fig:11}
\end{center}
\end{figure}

In Fig. \ref{fig:11}, we change
the parameters to $\sigma_g= 10, \sigma_h=1, \sigma_z=2,
\sigma_k=4$, $\gamma=10dB$ and $M=10$. In this case, channels between the relays and the eavesdropper
and between the relays and the primary-user are on average stronger than the channels between the relays and the destination. We note that
beamforming schemes can still attain good performance and we observe similar trends as before.

\begin{figure}
\begin{center}
\includegraphics[width = 0.5\textwidth]{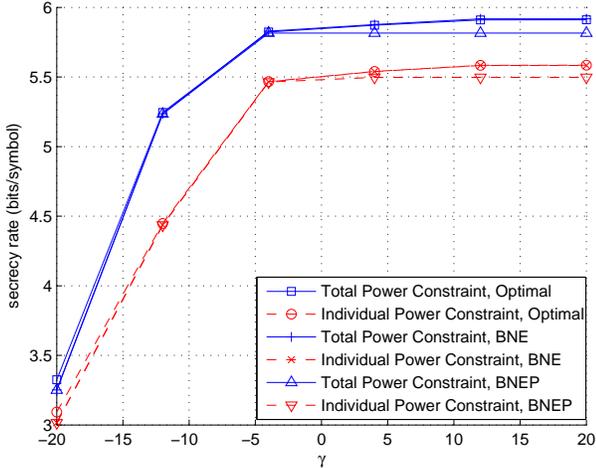}
\caption{AF secrecy rate vs. interference temperature $\gamma$. $
\sigma_g= 10, \sigma_h=2, \sigma_z=2, \sigma_k=4, M=10, P_s=P_T=0
dB$. } \label{fig:2}
\end{center}
\end{figure}

In Fig. \ref{fig:2}, we plot the optimal secrecy rate and the secrecy rates of the
two suboptimal null space beamforming schemes (under both total and individual power constraints) as a function of the
interference temperature limit $\gamma$. We assume that
$P_T=P_s=0 dB$. It is  observed that the
secrecy rate achieved by beamforming in the null space of both
the eavesdropper's and primary user's channels (BNEP) is almost
insensitive to different interference temperature limits when
$\gamma \geq -4 dB$ since it always forces the signal interference
to be zero regardless of the value of $\gamma$. It is further observed that
beamforming in the null space of the eavesdropper's channel (BNE) always
achieves near optimal performance regardless the value of   $\gamma$
under both total and individual power constraints.

\section{Conclusion}\label{sec:con}
In this paper, collaborative relay beamforming in
cognitive radio networks is studied under secrecy constraints. Optimal beamforming designs that maximize secrecy rates are investigated under both total and individual relay power
constraints. We have formulated the problem as a
semidefinite programming problem and provided an optimization framework.
In addition, we have proposed two sub-optimal null space beamforming schemes to simplify the computation. Finally, we have provided numerical results to illustrate the performances of different beamforming schemes.


\begin{thebibliography}{99}






\bibitem{Haykin}
S. Haykin ``Cognitive radio: brain-empowered wireless
communicatioins,'' \emph{IEEE J. Sel. Areas Commun}, vol.23, no.2,
pp.201-220, Feb 2005.

%
%
%
%
%

\bibitem{wyner}
A. Wyner  ``The  wire-tap channel,'' \emph{Bell. Syst Tech. J},
vol.54, no.8, pp.1355-1387, Jan 1975.
%
\bibitem{Csiszar}
I. Csiszar and J. Korner ``Broadcast channels with confidential
messages,'' \emph{IEEE Trans. Inform. Theory}, vol.IT-24, no.3,
pp.339-348, May 1978.

\bibitem{cheong}
S. K. Leung-Yan-Cheong and M. E. Hellman  ``The Gaussian wire-tap
channel,'' \emph{IEEE Trans. Inform. Theory}, vol.IT-24, no.4,
pp.451-456, July 1978.
%
%
%
%





%
\bibitem {Gan}
G. Zheng, K. Wong, A. Paulraj, and B. Ottersten,
``Collaborative-relay beamforming with perfect CSI: Optimum and
distributed implementation,'' \emph{IEEE Signal Process Letters},
vol. 16, no. 4, Apr. 2009

%
%
%
%
\bibitem{luo}
V. Nassab, S. Shahbazpanahi, A. Grami, and Z.-Q. Luo,`` Distributed
beamforming for relay networks based on second order statistics of
the channel state information,'' \emph{IEEE Trans. on Signal Proc.},
Vol. 56, No 9, pp. 4306-4316, Sept. 2008.
%
\bibitem{Gan1}
G. Zheng, K. K. Wong, A. Paulraj, and B. Ottersten, ``Robust
collaborative-relay beamforming,''  \emph{IEEE Trans. on Signal
Proc.}, vol. 57, no. 8, Aug. 2009
%
%

\bibitem{luo1}
Z-Q Luo , Wing-kin Ma , A.M.-C. So ,Yinyu Ye , Shuzhong Zhang
 ``Semidefinite relaxation of quadratic optimization problems''
\emph{IEEE Signal Proc. Magn.}, vol. 27, no. 3, May 2010

\bibitem{dong}
L. Dong, Z. Han, A. Petropulu and H. V. Poor, ``Secure wireless
communications via cooperation,'' \emph{ Proc. 46th Annual Allerton
Conf. Commun., Control, and Computing}, Monticello, IL, Sept. 2008.



\bibitem{jzhang} J. Zhang and M. C. Gursoy,
``Collaborative relay beamforming for secrecy,''  \emph{Proc. of the
IEEE International Conference on Communication (ICC)}, Cape Town,
South Africa, May 2010.

\bibitem{jzhang1}
J. Zhang and M.C. Gursoy,  ``Relay beamforming strategies for
physical-layer security,''  \emph{ Proc. of the 44th Annual
Conference on Information Sciences and Systems}, Princeton, March
2010

\bibitem{Pei}
Y. Pei, Y-C. Liang, L. Zhang, K. C. Teh,  and K. H. Li ``Secure
communication over MISO cognitive radio channel,'' \emph{IEEE
Trans. Wireless. Commun}, vol.9, no.4, pp.1494-1502, April. 2010.



\bibitem{matrix}
 G. Golub and C. F. Van Loan, Matrix Computations (3rd ed),  Johns Hopkins University Press,
1996

%
%
%

\bibitem{sedumi}
J. Sturm, ``Using SeDuMi 1.02: A MATLAB toolbox for optimization
over symmetric cones,'' \emph{Opt. Methods and Software}, vol.
11-12, pp. 625-653, 1999.

\bibitem{yalmip}
 J. Lofberg, ``YALMIP: A toolbox for
modeling and optimization in MATLAB,'' \emph{Proc. the CACSD Conf.},
Taipei, Taiwan, 2004.
%
\bibitem{boyd}
 S. Boyd and L.
Vandenberghe, Convex optimization. Cambridge, U.K.: Cambridge Univ.
Press, 2004.

\end{thebibliography}
\end{document}